\documentclass[aps,prb,twocolumn,showpacs,10pt,superscriptaddress,floatfix,longbibliography]{revtex4-1}

\usepackage{newtxtext,newtxmath}
\usepackage{graphicx}
\usepackage{subfigure}
\usepackage{amsmath}
\usepackage{amsfonts}
\usepackage{mathrsfs}
\usepackage{bbm}
\usepackage{subfigure}
\usepackage{xcolor}
\usepackage[colorlinks=true]{hyperref}
\usepackage{ulem}
\usepackage[T1]{fontenc}

\begin{document}

\title{Topological superconductivity with large Chern numbers in a ferromagnetic metal-superconductor heterostructure}	

\author{Yingwen Zhang}
\affiliation{School of Physics, Guangdong Provincial Key Laboratory of Magnetoelectric Physics and Devices, Sun Yat-sen University, Guangzhou 510275, China}
\affiliation{State Key Laboratory of Optoelectronic Materials and Technologies, Sun Yat-sen University, Guangzhou 510275, China}

\author{Dao-Xin Yao}
\email{yaodaox@mail.sysu.edu.cn}
\affiliation{School of Physics, Guangdong Provincial Key Laboratory of Magnetoelectric Physics and Devices, Sun Yat-sen University, Guangzhou 510275, China}
\affiliation{State Key Laboratory of Optoelectronic Materials and Technologies, Sun Yat-sen University, Guangzhou 510275, China}
\affiliation{International Quantum Academy, Shenzhen 518048, China}

\author{Zhi Wang}
\email{wangzh356@mail.sysu.edu.cn}
\affiliation{School of Physics, Guangdong Provincial Key Laboratory of Magnetoelectric Physics and Devices, Sun Yat-sen University, Guangzhou 510275, China}
 
\begin{abstract}
     Ferromagnetic metal-superconductor heterostructure with spin-orbit coupling is a promising candidate for topological superconductivity. Inspired by recent experimental progress in layered van der Waals metal-superconductor heterostructures, we study the interplay between the interface Rashba hopping and the intrinsic Dresselhaus spin-orbit coupling and demonstrate rich topological phases with five distinct Chern numbers.
     In particular, we find a topological state with the Chern number as large as four in a range of parameter space. We calculate the Berry curvatures that construct the Chern numbers and show that these Berry curvatures induce anomalous thermal Hall transport of the superconducting quasiparticles. We reveal chiral edge states in the topological phase and helical edge states in the topologically trivial phase, and show that the wave functions of these edge states mostly concentrate on the ferromagnetic metal layer of the heterostructure.

\end{abstract}
\maketitle
\section{Introduction}
Topological superconductors with Majorana modes have attracted much interest recently due to their possible applications in topological quantum computation\cite{kitaev2003fault,Kane2005QSHI, qi2011RMP, beenakker2013, stern2013topological,lutchyn2018review, aguado2017, Sato2017}. The topology of two-dimensional superconducting systems can be characterized by the Chern number\cite{read2000paired,Kane2005QSHI}, which is the summation of the momentum-space Berry curvature of the Bogoliubov–de Gennes Hamiltonian\cite{qi2010chiral,chiu2016classification,PhysRevLett.126.187001}. For superconductors with nonzero Chern number, the bulk-edge correspondence principle predicts Majorana chiral modes\cite{chiu2016classification}. Moreover, Majorana zero modes could exist in the presence of superconducting vortexes for superconductors with odd Chern numbers\cite{volovik1999fermion,read2000paired,qi2010chiral}. The pursuit of superconductors with nonzero Chern number is one of the central tasks in the present study of topological superconductivity\cite{nayak2008non,sarma2015majorana,flensberg2021engineered}.

Topological superconductors with nontrivial Chern numbers usually require a chiral superconducting gap functions\cite{read2000paired,tanaka2011symmetry,alicea2012new,Black_Schaffer2014}. The chiral superconducting gap was explored in a few materials\cite{luke1998time,mackenzie2003superconductivity,balatsky2006impurity,nelson2004odd,maeno2011evaluation,wang2018evidence,zhang2018observation,nandkishore2012chiral,isobe2018unconventional}. Recently, much effort is concentrated on achieving the effective chiral superconducting gap in the designed heterostructures where conventional s-wave superconductors are in closed contact with metallic systems such as topological insulators\cite{fu2008superconducting,wang2012coexistence,xu2014artificial,Sun2016Majorana,li2017origin,liu2018majorana} and spin-orbit coupling semiconductors\cite{sau2010twodimension,sau2010generic,sau2010non}. On the interface of these heterostructures, the delicate combination of the superconductivity, the spin-orbit coupling, and the Zeeman energy can bring topological superconducting states with nonzero Chern numbers\cite{alicea2012new,flensberg2021engineered}.

In these heterostructures, spin-orbit coupling is the central ingredient for topological superconductivity. Previous theoretical proposals for the required spin-orbit coupling fall into two distinct mechanisms. 
Firstly, there can be intrinsic spin-orbit coupling in the metal layer of the heterostructure, such as the surface states of the three-dimensional topological insulator and the spin-orbit coupling semiconductor.
Then a proper Zeeman energy can turn the system into the topological state \cite{sau2010twodimension,sato2010topological,fu2008superconducting}. An alternative proposal involves a heterostructure of a conventional superconductor and a half-metal\cite{eschrig2008triplet,chung2011topological,zhang2014topological}. The half-metallic layer has no intrinsic spin-orbit coupling. Instead, there is Rashba spin-orbit hopping between the half-metal and the conventional superconductor. The heterostructure will enter the topological superconducting phase when the half-metal has an odd number of Fermi surfaces.

\begin{figure}[b]
\centering
    \includegraphics[scale=0.30]{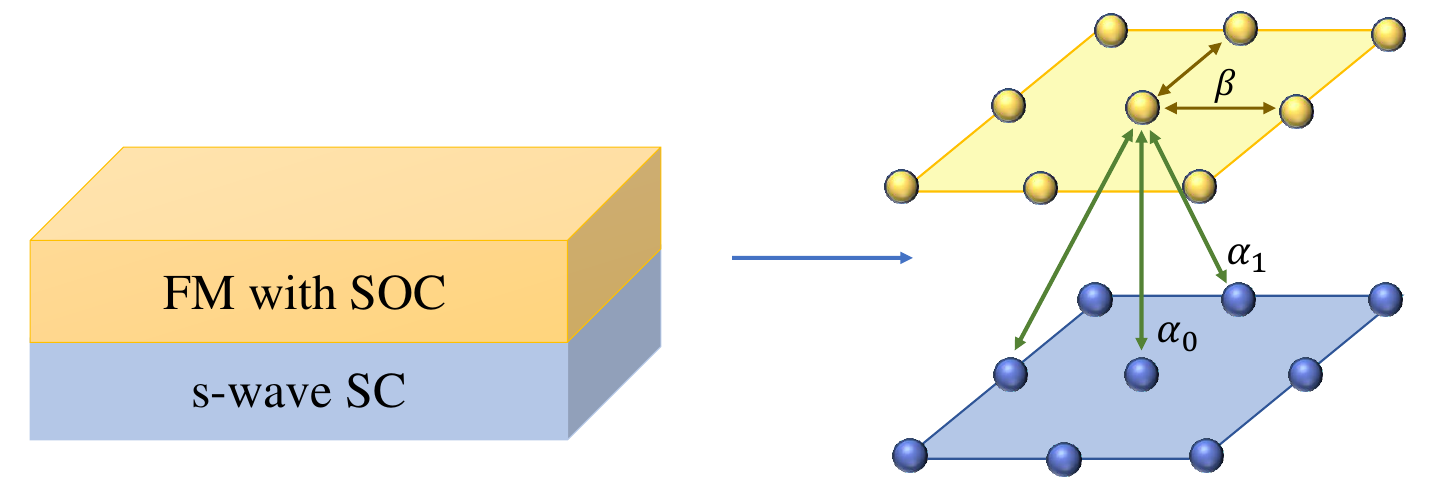}
    \caption{Schematics of the heterostructure constructed by ferromagnetic metal and s-wave superconductor. The ferromagnetic metal layer has intrinsic Dresselhaus spin-orbit coupling denoted by the tight-binding hopping parameter $\beta$. Meanwhile, there is an inter-layer Rashba spin-orbit hopping denoted by the tight-binding hopping parameter $\alpha_1$, while the spin-conserving inter-layer hopping is denoted by the tight-binding hopping parameter $\alpha_0$.}
    \label{Fig:Model}
\end{figure}

Recently, there has been considerable progress in the experimental realization of heterostructures between conventional superconductors and layered van der Waals (vdW) ferromagnetic materials\cite{huang2017layer,gong2017discovery,wang2018electric,fei2018two,deng2018gate}, such as the FGT/superconductor hybrid systems\cite{hu2023long}, $\rm{CrO_2}$/superconductor heterostructures\cite{keizer2006spin,zhang2020tunable} and $\rm{CrBr_3/NbSe_2}$ heterostructures\cite{kezilebieke2020topological,kezilebieke2021electronic,kezilebieke2022moire}. In these van der Waals heterostructures, a strong proximity effect can induce superconductivity in the ferromagnetic layer\cite{alidoust2015zero,halterman2018induced}, even in the presence of a significant Zeeman field. More interestingly, there can be the coexistence of intrinsic spin-orbit coupling in the ferromagnetic metal layer\cite{kim2018large,alghamdi2019highly} and extrinsic spin-orbit interlayer hopping between the ferromagnetic metal and superconductor\cite{chung2011topological}. Therefore, these ferromagnetic metal-superconductor heterostructures provide an interesting playground for investigating the interplay between the two distinct mechanisms for achieving topological superconductivity. In particular, one would expect rich topological phases due to the competition between the intrinsic Dresselhaus spin-orbit coupling in the ferromagnetic metal and the Rashba spin-orbit inter-layer hopping.

In this work, we study a two-layer model for the ferromagnetic metal-superconductor heterostructure as shown in Fig. \ref{Fig:Model}. The ferromagnetic metal layer has intrinsic Dresselhaus spin-orbit coupling and Zeeman energy, while the superconducting layer has an s-wave superconducting gap. The inter-layer hopping has both spin-conserving and spin-flipping components. The spin-conserving hopping is the conventional electron tunneling between two layers while the spin-flipping hopping is the Rashba spin-orbit hopping due to the inversion symmetry breaking on the interface between the two layers. We calculate the phase diagrams of the system with typical parameters and show topological phases with five distinct Chern numbers. 
In particular, we find a topological phase with the Chern number as large as four when both the superconducting layer and the metallic layer are nearly half filled.
We calculate the Berry curvatures that construct the Chern numbers and show that the Berry curvatures induce anomalous thermal Hall effect for the superconducting quasiparticles.
Finally, we study the boundary states and show chiral modes for the topological phases. Interestingly, we also find helical edge modes even in the topologically trivial phase. We find that these helical modes are protected by a hidden chiral symmetry of the Hamiltonian. We exhibit the wave functions of the edge modes and find that they are concentrated on the ferromagnetic metal layer of the system.

The remainder of this paper is organized as follows. In Sec. II, we introduce the ferromagnetic metal and s-wave superconductor two-layer model that is considered in the work, and we show the phase diagram which exhibits topological states with different Chern numbers. In Sec. III, we show the Berry curvatures in the momentum space which construct the Chern numbers and the anomalous thermal Hall conductivity governed by these Berry curvatures. In Sec. IV, we demonstrate the dispersion of the edge states with open boundary conditions in one direction and show the real-space distribution of the edge state wave functions. Finally, we give a summary in Sec. V.

\section{Model and phase diagram}
In the designs of realistic topological superconducting systems, one of the main theoretical proposals is the heterostructure of conventional superconductors and spin-orbit coupling semiconductors. In these designs, the central ingredient is the spin-orbit coupling. To be specific, there are two different methods to incorporate spin-orbit coupling into the system. One proposal is to consider a semiconductor with intrinsic spin-orbit coupling \cite{fu2008TSCvortex,sau2010twodimension,sato2010topological}, then the direct proximity of the superconducting pairing will induce topological superconductivity with proper Zeeman field. Another proposal involves a half-metal without intrinsic spin-orbit coupling. In contrast, the proximity of the superconducting pairing must include a Rashba spin-orbit inter-layer hopping which flips the spin of the Cooper pair\cite{chung2011topological}. Here, we consider a minimal model which takes account of both processes. This minimal model is a double-layer square-lattice tight-binding model, as shown in Fig. \ref{Fig:Model}. The ferromagnetic metal layer is described by the Hamiltonian
\begin{eqnarray}
H_{\textrm {FM}} = &&-t_1\sum_{\langle ij \rangle\alpha}c^{\dagger}_{i\alpha}c_{j\alpha}+\mu_1 \sum_{i\alpha}c^{\dagger}_{i\alpha}c_{i\alpha}\\
&&+i\beta \sum_{\langle ij \rangle\alpha\gamma}c^{\dagger}_{i\alpha}(\boldsymbol{\sigma}_{\alpha\gamma} \times \boldsymbol{d}_{ij})_zc_{j\gamma}+M_z\sum_{i}(c^{\dagger}_{i\uparrow}c_{i\uparrow}-c^{\dagger}_{i\downarrow}c_{i \downarrow}), \nonumber
    \label{eq:1}
  \end{eqnarray}
where $\langle ij \rangle$ represents the nearest-neighboring sites, $t_1$ is the nearest-neighbor hopping energy, $\alpha$ and $\gamma$ represent the spin index, $\mu_1$ is the chemical potential, $M_z$ is the effective Zeeman energy, $\beta$ represents to the strength of Dresselhaus spin-orbit coupling, $\boldsymbol{\sigma}_{\alpha\gamma}$ corresponds to Pauli matrices in spin space, and $\boldsymbol{d}_{ij}$ is a unit vector which can be written as $(-(i_x-j_x), i_y-j_y, 0)$. The lattice constants are set to unity. 

The s-wave superconducting layer is described by the Hamiltonian
\begin{equation}
    \begin{aligned}
     H_{\textrm {SC}}= -t_2\sum_{\langle ij \rangle\alpha}f^{\dagger}_{i\alpha}f_{j\alpha}+\mu_2 \sum_{i\alpha}f^{\dagger}_{i\alpha}f_{i\alpha}+\Delta\sum_i f^{\dagger}_{i\uparrow}f^{\dagger}_{i\downarrow},
    \end{aligned}\label{eq:2}
  \end{equation}
where $t_2$ is the nearest-neighbor hopping energy, $\mu_2$ is the chemical potential, and $\Delta$ represents the superconducting order parameter.

The two layers are coupled through  electron hopping and the Hamiltonian can be written as
\begin{equation}
    \begin{aligned}    
    H_{\textrm {I}}={\alpha}_0\sum_{i\alpha}f^{\dagger}_{i\alpha}c_{i\alpha}+i\alpha_1 \sum_{\langle ij \rangle\alpha\gamma}c^{\dagger}_{i\alpha}(\boldsymbol{\sigma}_{\alpha\gamma} \times \boldsymbol{d}^{\prime}_{ij})_zf_{j\gamma},
    \end{aligned}\label{eq:3}
  \end{equation}
where $\alpha_0$ and $\alpha_1$ represent the hopping parameters for spin-flip hopping and spin-conserving hopping when electrons hopping across the interlayer, respectively, $\boldsymbol{d}^{\prime}_{ij}$ can be written as $(i_x-j_x, i_y-j_y, 0)$, is the unit projection vector from j to i in $x-y$ plane. This vector represents an electron hopping that resembles the form of Rashba spin-orbit coupling.

The total Hamiltonian for the model in real space can be written as $H=H_{\rm{FM}}+H_{\rm{SC}}+H_{\rm{I}}$. It can be transformed into the momentum space when we consider periodic boundary conditions. In the presence of superconductivity, it is more convenient to formulate the total Hamiltonian in the Bogoliubov–de Gennes (BdG) form, which is written as
\begin{equation}
H=\frac{1}{2}\sum_{\boldsymbol{k}} \psi_{\boldsymbol{k}}^{\dagger}H_{\rm{BdG}}(\boldsymbol{k})\psi_{\boldsymbol{k}},
\label{eq:4}
\end{equation}	
where we define the Nambu spinor operator $\psi_{\boldsymbol{k}}^{\dagger}=(c_{{\boldsymbol{k}}\uparrow}^{\dagger},c_{{\boldsymbol{k}}\downarrow}^{\dagger},f_{{\boldsymbol{k}}\uparrow}^{\dagger},f_{{\boldsymbol{k}}\downarrow}^{\dagger},c_{-{\boldsymbol{k}}\uparrow},c_{-{\boldsymbol{k}}\downarrow},f_{-{\boldsymbol{k}}\uparrow},f_{-{\boldsymbol{k}}\downarrow})$, and the mean-field BdG Hamiltonian is written as
\begin{equation}
\begin{aligned}
H_{\rm{BdG}}(\boldsymbol{k}) &= \frac{1}{2}\epsilon _1(\boldsymbol{k})\tau_z(s_0+s_z)\sigma_0+\frac{1}{2}\epsilon _2(\boldsymbol{k})\tau_z(s_0-s_z)\sigma_0\\
&+\frac{1}{2}\beta \sin k_x \tau_z(s_0+s_z)\sigma_y + \frac{1}{2}\beta \sin k_y \tau_0(s_0+s_z)\sigma_x  \\ 
&+\alpha_0 \tau_zs_x\sigma_0-\alpha_1 \sin k_x\tau_zs_x\sigma_y +\alpha_1 \sin k_y\tau_0s_x\sigma_x \\
&+\frac{1}{2}M_z\tau_z(s_0+s_z)\sigma_z-\frac{1}{2}\Delta \tau_y(s_0-s_z)\sigma_y,
\end{aligned}\label{eq:5}
\end{equation}
where $\epsilon _1(\boldsymbol{k})=-2t_1(\cos k_x+\cos k_y)+\mu_1$, $\epsilon _2(\boldsymbol{k})=-2t_2(\cos k_x+\cos k_y)+\mu_2$, and $\boldsymbol{\tau}, \boldsymbol{s}, \boldsymbol{\sigma}$  are Pauli matrices in the particle-hole, layer, and spin degrees of freedom, respectively.

The topology of this two-layer system is characterized by the BdG Hamiltonian $H_{\rm BdG}(\boldsymbol{k})$, which is effectively a single particle Hamiltonian with particle-hole symmetry. The system has a Zeeman energy which breaks the time-reversal symmetry. As a result, the chiral symmetry of the system is also broken, and the system belongs to the D class \cite{schnyder2008classification,ryu2010topological,chiu2016classification} in the topological classification. The topological number is the Chern number which is the summation of the Berry curvatures in the momentum space
\begin{equation}
\begin{aligned}
\Omega_{\mu\upsilon}^n(\boldsymbol{R})=i\sum_{m\neq n}\frac{\left \langle m|\partial H/\partial R^{\mu}|n \right \rangle \left \langle n|\partial H/\partial R^{\upsilon}|m \right \rangle}{(E_m-E_n)^2}-(\mu \leftrightarrow \upsilon).
\end{aligned}\label{eq:6}
\end{equation}
The summation of the Berry curvatures of bands with negative energies provides the Chern number of the system. 

\begin{figure}[t]
    \includegraphics[scale=0.38]{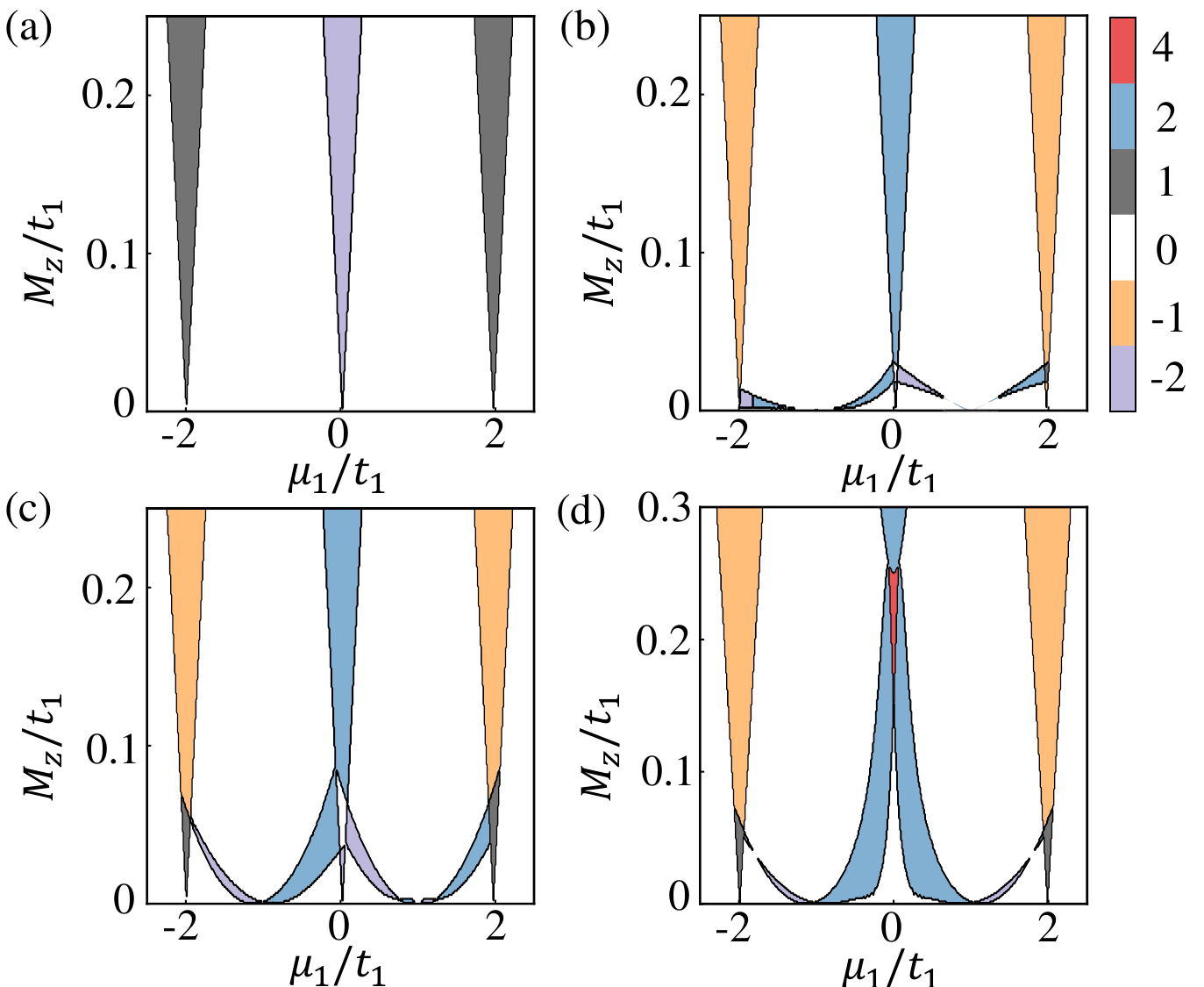}
    \caption{Topological phase diagrams with distinct Chern numbers for four typical parameters. (a) the intrinsic Dresselhaus spin-orbit coupling dominating regime with $\alpha_1/t_1=1/160$ and $\beta/t_1=1/8$. (b) the balanced regime with $\alpha_1/t_1=1/8, \beta/t_1=1/16$. (c) the inter-layer Rashba spin-orbit hopping dominating regime with $\alpha_1/t_1=1/8, \beta/t_1=1/160$. Other parameters are taken as $\Delta/t_2=1/10$, $\alpha_0/t_1=1/4$, and ${\mu}_2/t_2=1$. (d) the near half-filled regime with ${\mu}_2=0$ while other parameters are taken the same as (b).}
    \label{Fig:Phase}
\end{figure}

 We calculate the Chern number of the BdG Hamiltonian and find that this two-layer system exhibits multiple topological phases with different Chern numbers. This complicated phase diagram comes from the competition between the intra-layer Dresselhaus spin-orbit coupling and the inter-layer Rashba spin-orbit hopping. To reveal this competition, we demonstrate phase diagrams for the Chern number with four typical parameters in Fig.~\ref{Fig:Phase}. We first demonstrate the scenario where the Dresselhaus spin-orbit coupling dominates the Rashba spin-orbit inter-layer hopping. As shown in Fig.~\ref{Fig:Phase}(a), we find a phase diagram that qualitatively replicates the phase diagram of the single-layer models, which was studied in previous works\cite{sau2010twodimension,sato2010topological,sato2010non}. There are three topological regions with two distinct Chern numbers of one and minus two. The phase boundary also closely resembles the single-layer system. The gap of $H_{\rm{BdG}}(\boldsymbol{k})$ closes at momenta of $\Gamma$, $X(0,\pi),(\pi,0)$, $M(\pi,\pi)$ points. For the case of gap closing at $\Gamma$ point, the intrinsic Dresselhaus spin-orbit coupling and inter-layer Rashba spin-orbit hopping in $H_{\rm{BdG}}$ are negligible. Then the secular equation for the gap closing condition $\det[H_{\rm{BdG}}(\Gamma)]=0$ is analytically solvable, and we can write down the phase transition between the topologically trivial phase and the topological phase with Chern number equals to one (the case when the Fermi surface is at the bottom of the ferromagnet metal' energy band) as
\begin{equation}
\begin{aligned}
M_{z1}=\sqrt{(a_1{\mu}_1^2+a_2{\mu}_1+a_3)/a_1},
\end{aligned}\label{eq:7}
\end{equation}  
where $a_1=16t_2^2-8t_2{\mu}_2+{\mu}_2^2+{\Delta}^2$, $a_2=-2\alpha_0^2({\mu}_2-4t_2)-8t_1a_1$, and $a_3=\alpha_0^4-8\alpha_0^2t_1(4t_2-{\mu}_2)+16t_1^2a_1$. Similarly, we can also solve the secular equation to obtain the phase boundary between the topologically trivial phase and the topological phase with the $C=-2$ (the gap closes at X points) as
\begin{equation}
\begin{aligned}
M_{z2}=\sqrt{(b_1{\mu}_1^2+b_2{\mu}_1+b_3)/b_1},
\end{aligned}\label{eq:8}
\end{equation}	
where $b_1={\mu}_2^2+{\Delta}^2$, $b_2=-2\alpha_0^2{\mu}_2$, and $b_3=\alpha_0^4$. We notice that these expressions resemble the phase boundaries for the single-layer system\cite{sau2010twodimension}, although the parameters become much more complicated due to the two-layer nature of the model. The resemblance indicates that the two-layer nature of the model does not bring much complexity because the dominant ingredient is the intrinsic Dresselhaus spin-orbit coupling.
In Fig. \ref{Fig:Phase}(b), we demonstrate the scenario where the Rashba spin-orbit inter-layer hopping dominates the Dresselhaus spin-orbit coupling. In this scenario, the intrinsic spin-orbit coupling of the ferromagnetic metal is negligible and we simply come back to the half-metal/superconductor model that was introduced in Ref. [\onlinecite{chung2011topological}]. We find that the phase diagrams are similar to Fig. \ref{Fig:Phase}(a), except that the Chern numbers have a sign reversal.

\begin{figure*}[t]
    \includegraphics[scale=0.45]{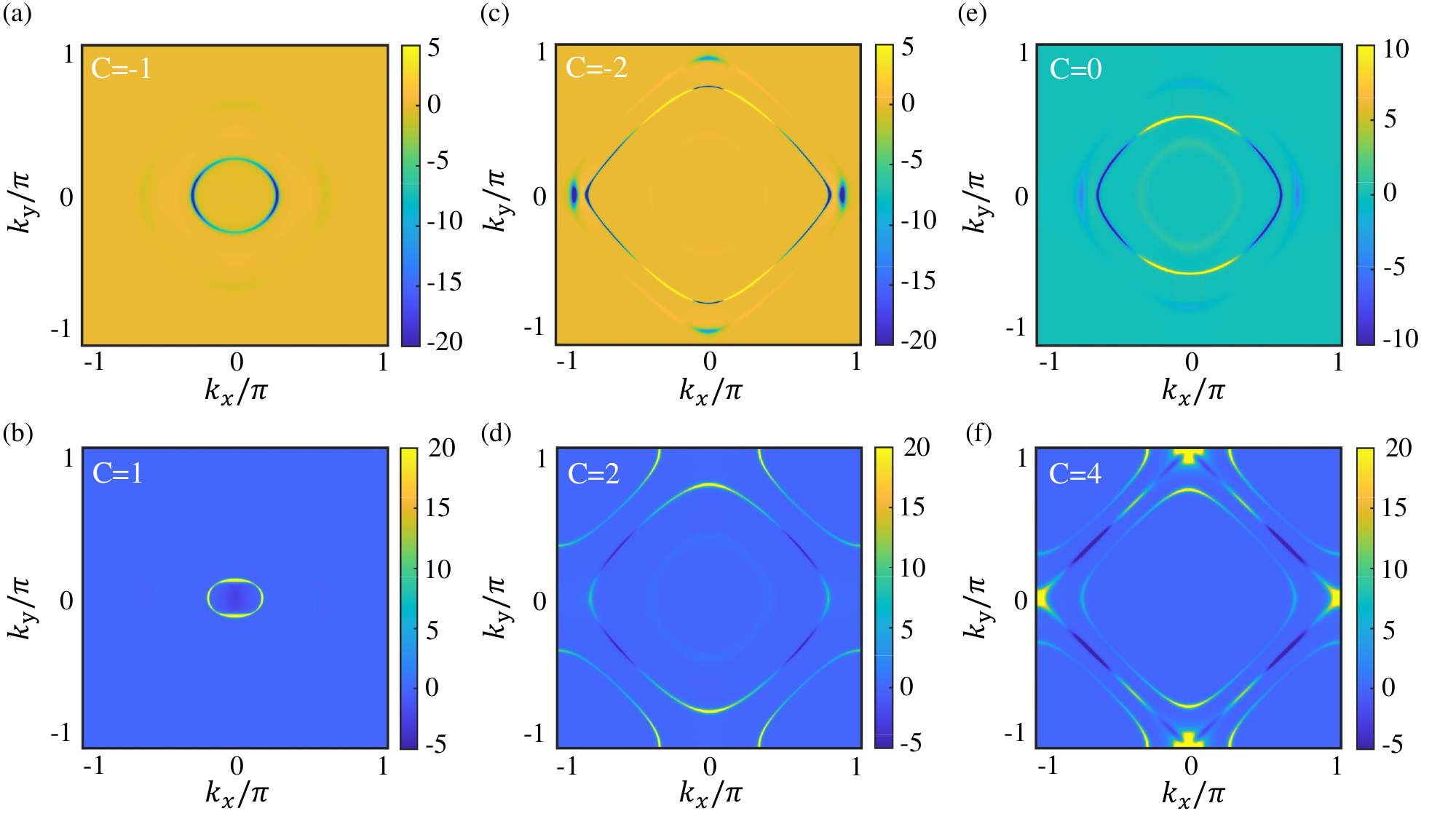}
    \caption{Momentum space Berry curvatures for typical parameters with six distinct Chern numbers. (a)-(b) Topological phase with $C=-1$ and $C=1$, with the Berry curvatures concentrate around the single Fermi surface of the ferromagnetic metal layer. (c-d) Topological phase with $C=-2$ and $C=2$, with the Berry curvatures concentrated around the two Fermi surfaces of the ferromagnetic metal layer. (e) Topologically trivial phase with $C=0$, with nonvanishing Berry curvatures around the Fermi surfaces of the ferromagnetic layer. (f) Topological phase with $C=4$, with both the superconducting layer and the ferromagnetic metal layer are nearly half filled.}
    \label{Fig:Berry}
\end{figure*}

Now we come to the more complex and interesting scenario where the Dresselhaus spin-orbit coupling and the Rashba spin-orbit inter-layer hopping are comparable. In this case, the phase diagram exhibits complicated topological ordering as shown in Fig. \ref{Fig:Phase}(c). There are a number of phase regions with distinct Chern numbers of plus/minus one, plus/minus two, and zero. Looking at the boundary of the phase transition, we find that the phase transition between topological states and trivial states as shown in Eqs. (\ref{eq:7}) and (\ref{eq:8}) still exist. However, there are additional phase transitions that enrich the phase diagram. In particular, we find phase transitions between the topological states of $C=1$ and $C=-1$. This phase transition comes from the competition between the Dresselhaus spin-orbit coupling and the Rashba spin-orbit inter-layer hopping. Let us reveal this competition by looking closely at the single-particle Hamiltonian of the two-layer system
\begin{equation}       %开始数学环境
H = \left(                 %左括号
  \begin{array}{cc}   %该矩阵一共3列，每一列都居中放置
    (\epsilon _1(\boldsymbol{k})\sigma_0+M_z\sigma_z+\beta(\sin k_y\sigma_x+\sin k_x\sigma_y) &  c.p.\\  %第一行元素
    \alpha_0\sigma_0+\alpha_1(\sin k_y\sigma_x+\sin k_x\sigma_y) &\epsilon _2(\boldsymbol{k})\sigma_0 \\  %第二行元素
  \end{array}\label{eq:9}
\right).              %右括号
\end{equation}
This is a $4 \times 4$ matrix where the upper left $2 \times 2$ blocks describe the ferromagnetic metal and the lower right $2 \times 2$ block describes the conventional superconductor.
In this $4 \times 4$ matrix, we notice that the Dresselhaus spin-orbit coupling and the Rashba spin-orbit inter-layer hopping stay at different $2 \times 2$ blocks. However, they can be put together by diagonalizing the ferromagnetic metal block with a unitary transformation,
\begin{eqnarray}       %开始数学环境
U=\left(                 %左括号
  \begin{array}{cc}   %该矩阵一共3列，每一列都居中放置
    e^{\frac{i\theta}{2}\hat{\bf n}\cdot \hat{\bf \sigma}} &  0\\  %第一行元素
    0& \sigma_0 \\  %第二行元素
  \end{array}\label{eq:10}
\right)    ,             %右括号
\end{eqnarray}
where $\theta=\arctan (\beta\sqrt{(\sin k_x^2+\sin k_y^2)}/M_z)$, and $\hat{\bf n}=(-\sin k_x,\sin k_y,0)/\sqrt{\sin k_x^2+\sin k_y^2}$ is a unit vector. After this unitary transformation, the Hamiltonian is written as
\begin{equation}       %开始数学环境
H = \left(                 %左括号
  \begin{array}{cc}   %该矩阵一共3列，每一列都居中放置
    \epsilon _1(\boldsymbol{k})\sigma_0+\frac{M_z}{\cos \theta}\sigma_z & h_{12}\\  %第一行元素
    h_{21}& \epsilon _2(\boldsymbol{k}) \\  %第二行元素
  \end{array}\label{eq:11}
\right),                 %右括号
\end{equation}
where 
 \begin{equation}
\begin{aligned}
h_{12}
&=(\alpha_0\cos \frac{\theta}{2}-2i\alpha_1\sin \frac{\theta}{2}\sin k_x \sin k_y/\sqrt{\sin k_x^2+\sin k_y^2})\sigma_0\\
&+(\alpha_1\cos \frac{\theta}{2}\sin k_y-i\alpha_0sin\frac{\theta}{2}\sin k_x/\sqrt{\sin k_x^2+\sin k_y^2})\sigma_x\\
&-(\alpha_1\cos \frac{\theta}{2}\sin k_x-i\alpha_0sin\frac{\theta}{2}\sin k_y/\sqrt{\sin k_x^2+\sin k_y^2})\sigma_y \\
&-(\alpha_1\sin \frac{\theta}{2}(\sin ^2k_x-\sin ^2k_y)/\sqrt{\sin k_x^2+\sin k_y^2})\sigma_z.
\end{aligned} \label{eq:12}
\end{equation}	
We notice that both Dresselhaus spin-orbit coupling and the Rashba spin-orbit inter-layer hopping are now transformed to the off-diagonal $2\times 2$ block of the Hamiltonian. The model is now equivalent to the half-metal/superconductor system introduced by Ref. [\onlinecite{chung2011topological}], where the spin-orbit inter-layer hopping is a combination of the Dresselhaus and the Rashba terms. By simplifying the $h_{12}$ in Eq. (\ref{eq:11}), we can get the real and imaginary terms of the upper right part in $h_{12}$, respectively,

 \begin{equation}
\begin{aligned}
&Re (h_{12})=(\alpha_1\cos \frac{\theta}{2}+\alpha_0\sin \frac{\theta}{2}/\sqrt{\sin k_x^2+\sin k_y^2})\sin k_y \\
&Im (h_{12})=(\alpha_1\cos \frac{\theta}{2}-\alpha_0\sin \frac{\theta}{2}/\sqrt{\sin k_x^2+\sin k_y^2})\sin k_x. \\
\end{aligned}\label{eq:13}
\end{equation}  
Now it is clear that the off-diagonal elements of the $h_{12}$ term are topologically equivalent to $\sin k_y+i\sin k_x$ or $\sin k_y-i\sin k_x$, depending on the relative sign of the two coefficients in Eq. (\ref{eq:13}). For two distinct cases, the system corresponds to effective $p+ip$ and $p-ip$ chiral topological superconductivity, and their corresponding Chern numbers are plus and minus one, respectively. The phase transition condition is the zero for the imaginary part which is written as 
\begin{eqnarray}
tan\frac{\theta}{2}=\frac{\alpha_1}{\alpha_0}\sqrt{\sin k_x^2+\sin k_y^2}.
\end{eqnarray}
This function provides a rough analytical understanding of the phase boundaries between $C=1$ and $C=-1$ in Fig. \ref{Fig:Phase}(c), which can be understood as the exact balance between the intra-layer Dresselhaus spin-orbit coupling and the inter-layer Rashba spin-orbit hopping.

Finally, we demonstrate topological phases with a Chern number as large as four, as shown in Fig. \ref{Fig:Phase}(d). This topological phase with a large Chern number requires a delicate balancing of the parameters. For example, the Fermi surfaces of both the ferromagnetic metal and the superconductor have to overlap near the nesting position. This large Chern number is a result of the two-layer nature of the model. As we will show, it requires all four bands of the system to be topological. Therefore, it does not appear in the previous one-layer models\cite{sau2010twodimension}.

\begin{figure}[h]
\centering
    \includegraphics[scale=0.38]{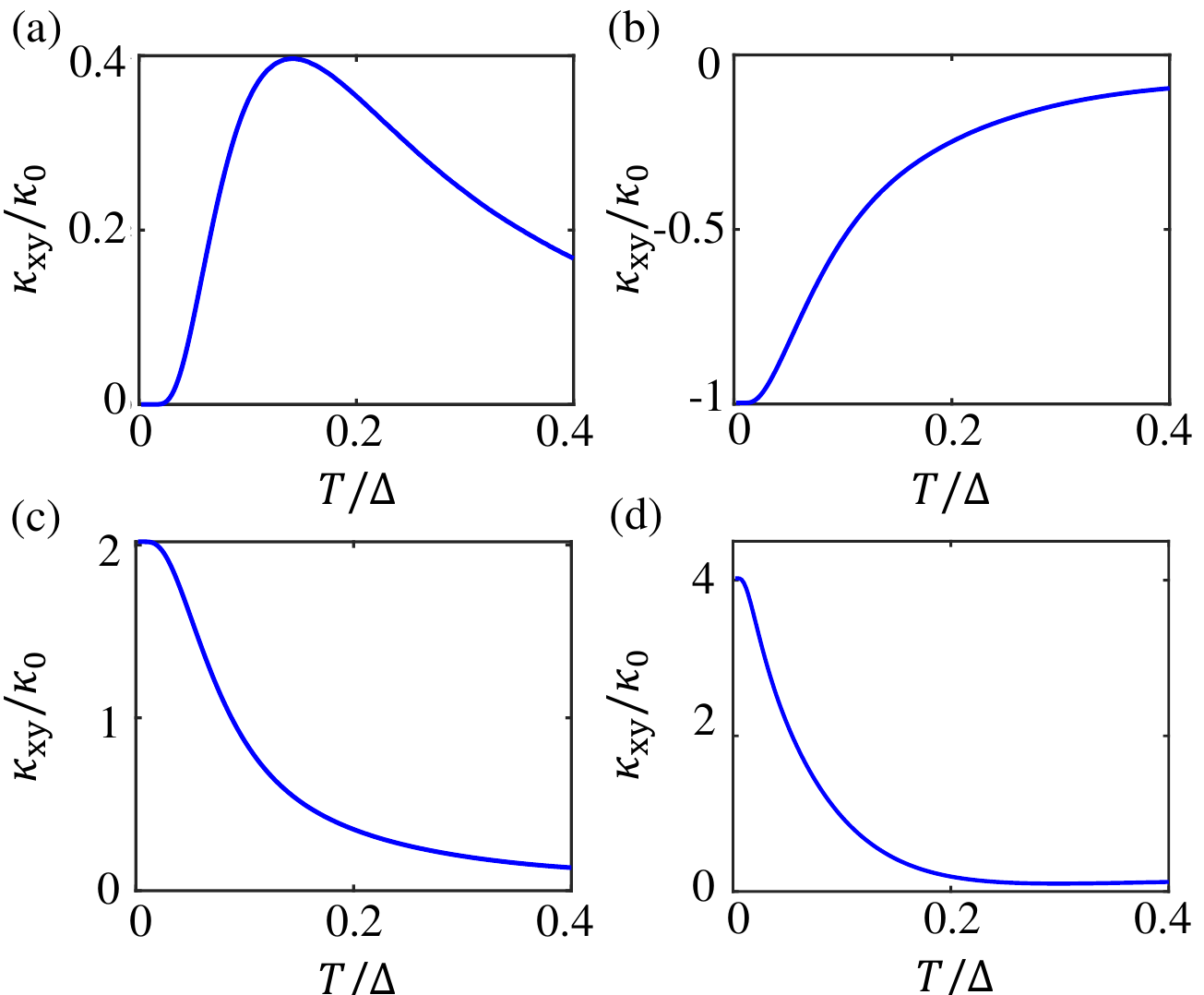}
    \caption{Anomalous thermal Hall conductivity as a function of temperature for four typical states with different Chern numbers. (a) Topologically trivial state with $C=0$. The zero temperature limit of the anomalous thermal Hall conductance is zero. (b-d) Topological states with $C=-1,2,4$. The zero temperature limit of the anomalous thermal Hall conductance is $C\kappa_0$ with $\kappa_0 = \pi k_B^2T/6\hbar$.}
    \label{Fig:ThermalHall}
\end{figure}

\section{Berry curvature}
The topological phase diagram with different Chern numbers can be understood more clearly by checking the Berry curvatures in the momentum space.
For this purpose, we illustrate Berry curvatures for typical Chern numbers in Fig. \ref{Fig:Berry}. We first look at the Berry curvatures with $C= \pm 1$, where one Majorana chiral mode is expected at the edge, and Majorana zero mode would appear in the presence of a superconducting vortex. As shown in Fig. \ref{Fig:Berry}(a) and Fig. \ref{Fig:Berry}(b), The Berry curvatures peak at a circle in the Brillouin zone. In fact, these circles are exactly the Fermi surface of the ferromagnetic metal. Due to the relatively large Zeeman energy, the two bands of the ferromagnetic metal split, and only one of them intersects with the Fermi energy. 
The Dresselhaus spin-orbit coupling or the Rashba spin-orbit inter-layer hopping modulates the Bogoliubov eigenstates around the Fermi surfaces and induces Berry curvatures that sum to the Chern number $C= \pm 1$.
In Fig. \ref{Fig:Berry}(c) and Fig. \ref{Fig:Berry}(d), we show the Berry curvatures corresponding to the Chern numbers $C=\pm 2$. In these cases, the chemical potential is lifted so that it touches both the energy bands of the ferromagnetic metal, even though they energetically split by the Zeeman energy. In this regime, the Berry curvatures also concentrate around the two Fermi surfaces of the ferromagnetic metal. We notice that the Fermi surfaces of the ferromagnetic metal are near nesting, which resembles the results of the one-layer model. While roughly one Fermi surface contributes a Chern number of plus/minus one, the total summation of the Berry curvatures provides a Chern number of plus/minus two. 
The Berry curvatures of the topologically trivial state with $C=0$ are shown in Fig. \ref{Fig:Berry}(e). It is clear that the Berry curvatures are non-vanishing while the summation is zero.
Finally, we examine the topological states with the large Chern number of $C=4$. As shown in Fig. \ref{Fig:Berry}(f), the Berry curvatures peak at the four Fermi surfaces of the ferromagnetic metal and the superconductor. These four Fermi surfaces are all near nesting, which causes complicated inter-band coupling effects. We believe that this near-nesting Fermi surface is the key ingredient to achieving the large Chern numbers in this system.
\begin{figure*}[t]
    \includegraphics[scale=0.6]{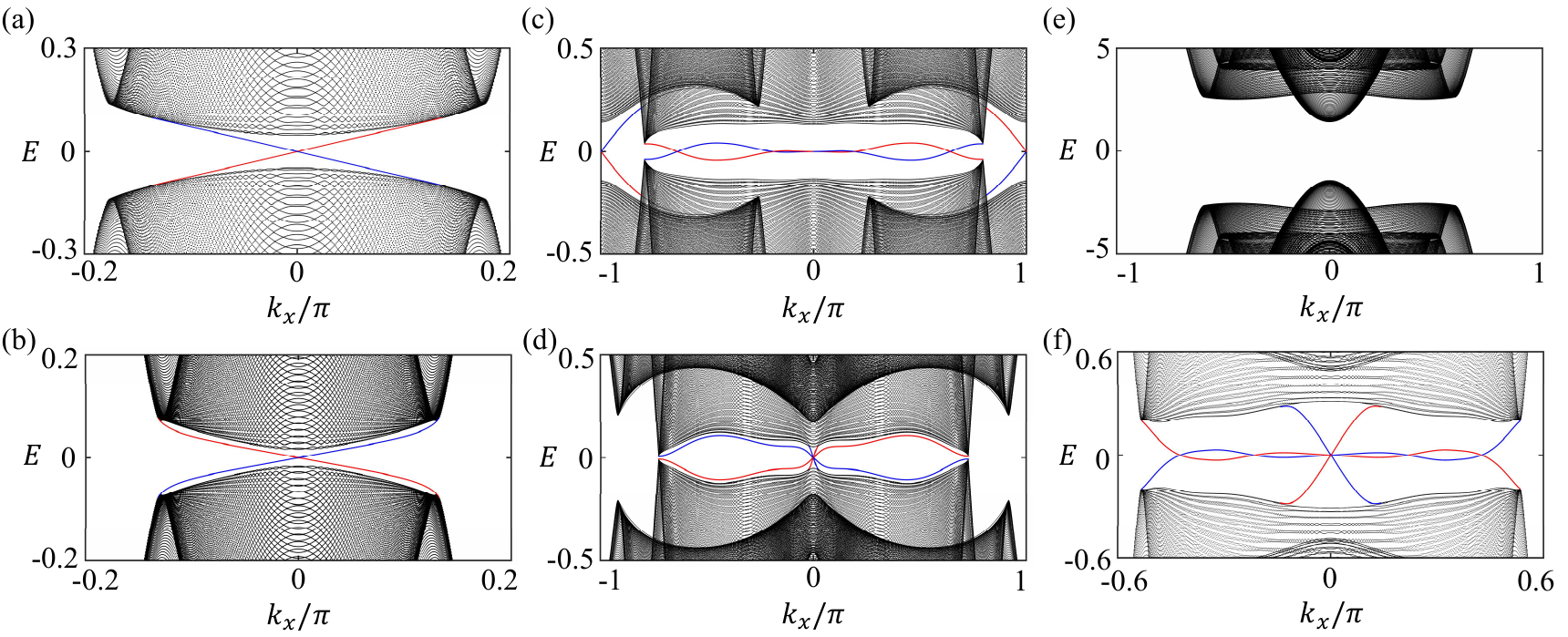}
    \caption{Energy spectra of the system as a function of the x-direction momentum, with open boundary condition in the y-direction. The blue (red) line represents states on the left (right) edge. (a)-(b) topological states with odd Chern numbers $C=-1$ and $C=1$. One majorana chiral mode propagates along opposite direction at different edge in these two situations. (c)-(d) topological states with Chern numbers $C=2$ and $C=-2$. Two majorana chiral modes propagate along opposite direction at different edge in these two situations. (e)-(f) topological trival states with Chern number $C=0$. There is no edge state in (e), but there are helical edge states in (f).}
    \label{Fig:edgestate}
\end{figure*}

The Berry curvatures are important to transport properties. The anomalous thermal Hall conductivity is related to the Berry curvatures through the formula\cite{smrcka1977transport,vafek2001quasiparticle,cvetkovic2015berry,PhysRevLett.126.187001},
 \begin{equation}
\begin{aligned}
\kappa_{xy}=\frac{1}{T}\sum_n\int[d\textbf{k}](\Omega_{n\textbf{k}})\int_{E_{n\textbf{k}}}^{\infty}d\eta\eta^2f'(\eta,T),
\end{aligned}\label{eq:14},
\end{equation}
where $f(E_{n\textbf{k}},T)=1/(e^{E_{n\textbf{k}}/T}+1)$ is the Fermi-Dirac distribution at temperature $T$, $f'$ is its derivative with respect to $E_{n\textbf{k}}$, $E_{n\textbf{k}}$ and $\Omega_{n\textbf{k}}$ are energy and the Berry curvature in momentum space with the energy band index $n$, respectively.
We show the temperature dependence of anomalous thermal Hall conductivity with different Chern numbers in Fig. \ref {Fig:ThermalHall}.
We find that the low-temperature limit of the thermal Hall conductivity is determined by the Chern number and the universal value of $\kappa_0=\pi k_B^2T/6\hbar$. In the high-temperature limit, the thermal Hall conductivity gradually decays to zero. We particularly note that the thermal Hall conductivity is non-vanishing even for zero Chern numbers. As shown in Fig. \ref {Fig:ThermalHall}(a), the thermal Hall conductivity is actually in the same order as those for nonzero Chern numbers at finite temperatures. The reason for the significant thermal Hall conductivity in the topologically trivial regime is well understood from the Berry curvature distribution shown in Fig. \ref{Fig:Berry}(e). The Berry curvatures are comparable to the topologically nontrivial regions even though their summation gives a zero Chern number. These transport signals are experimentally measurable and would provide information for both the Chern number and the Berry curvatures of topological superconducting systems.

\section{Chiral Majorana Edge states}
For open boundary systems, the bulk-edge correspondence predicts the chiral Majorana modes in the class D topological states with nonzero Chern numbers \cite{PhysRevB.82.184516, PhysRevB.92.064520}. To explicitly demonstrate the chiral Majorana edge modes, we consider the two-dimensional lattice with periodic boundary condition in the x direction and open boundary condition along the y direction. With this boundary condition, the momentum in the x direction is still a valid quantum number.
We numerically solve the Bogoliubov–de Gennes equation for this open boundary system, and show the energy spectra as a function of the momentum in the x direction, as shown in Fig. \ref{Fig:edgestate}.
For the topological states with odd Chern numbers of $C=\pm1$, we find one Majorana edge mode which propagates along the opposite direction at the different edges. 
In Fig. \ref {Fig:edgestate}(a), the blue edge localizes at the left boundary moving downwards with velocity $v_y<0$, while the red edge state localizes at the right boundary moving upwards with velocity $v_y>0$. In Fig. \ref {Fig:edgestate}(b), the chirality of the Majorana modes flips when the Chern number changes sign.

For larger Chern numbers of $C = 2$, we find two chiral Majorana edge states propagating in the same direction, as shown in Fig. \ref {Fig:edgestate}(c). These two chiral Majorana modes are energetically split, therefore can not be simply understood as part of a single chiral Dirac mode. In Fig. \ref {Fig:edgestate}(d), we show the edge states for $C=-2$, it is obvious that the propagation direction of edge states is reversed, consistent with expectation from the Chern number.

Finally, we analyze the results of topological trivial states. For Chern number equals zero, the simple analysis would expect the absence of edge states. This is indeed the case for most topologically trivial states as shown in Fig. \ref {Fig:edgestate}(e). However, in the numerical solutions, we find unexpected edge states even in the topological trivial region. As shown in Fig. \ref {Fig:edgestate}(f), two chiral edge modes are propagating along opposite directions on the same edge. These edge states resemble the helical edge states that appeared in topological insulators. They are stable with simple parameter modifications of the model, suggesting that they are not simply accidental edge states. In fact, these edge states are topologically protected by a winding number. To reveal this winding number more explicitly, we examine the one-dimensional Hamiltonian $H(k_x)$ by setting $k_y=0$, 
\begin{equation}
\begin{aligned}
&H(k_x)=\frac{1}{2}\epsilon _1(k_x)\tau_z(s_0+s_z)\sigma_0+\frac{1}{2}\epsilon _2(k_x)\tau_z(s_0-s_z)\sigma_0\\
&+\frac{1}{2}\beta \sin k_x \tau_z(s_0+s_z)\sigma_y + \alpha_0 \tau_zs_x\sigma_0-\alpha_1 \sin k_x\tau_zs_x\sigma_y\\
&+\frac{1}{2}M_z\tau_z(s_0+s_z)\sigma_z-\frac{1}{2}\Delta \tau_y(s_0-s_z)\sigma_y,
\end{aligned}\label{eq:151}
\end{equation}
This one-dimensional Hamiltonian has both particle-hole symmetry and chiral symmetry. We can define particle-hole operator $\hat{P}=\tau_x s_0 \sigma_0 \hat{\kappa}$, where $\hat{\kappa}$ is the complex conjugation operator, and the particle-hole symmetry is explicitly written as $H(k_x)=-\hat{P}H(-k_x)\hat{P}^{\dagger}$. The chiral operator is defined as $\hat{C}=\tau_x s_0 \sigma_0$, and the Chiral symmetry is written as $H(k_x)=-\hat{C}H(k_x)\hat{C}^{-1}$. The combination of these two symmetry operations gives $H(k_x)=H^{\top}(-k_x)$, which suggests that the one dimensional Hamiltonian
$H(k_x)$ belongs to class BDI \cite{ryu2010topological} in the tenfold way of Altland-Zirnbauer classification. Then the winding number can be defined as\cite{sato2011topology,ii2012theory}
 \begin{equation}
\begin{aligned}
w=\frac{1}{4\pi i}\int_{-\pi}^{\pi}dk_x Tr[\hat{C}H^{-1}\partial_{k_x}H].
\end{aligned}\label{eq:16}
\end{equation}
We calculate the winding number for Fig. \ref{Fig:edgestate}(f), and find that $w = 2$ which is consistent with the number of edge states. Similarly, we can calculate the one-dimensional Hamiltonian $H(k_y)$ when studying the open boundary condition in the x-direction.

\begin{figure}[t]
\centering
    \includegraphics[scale=0.36]{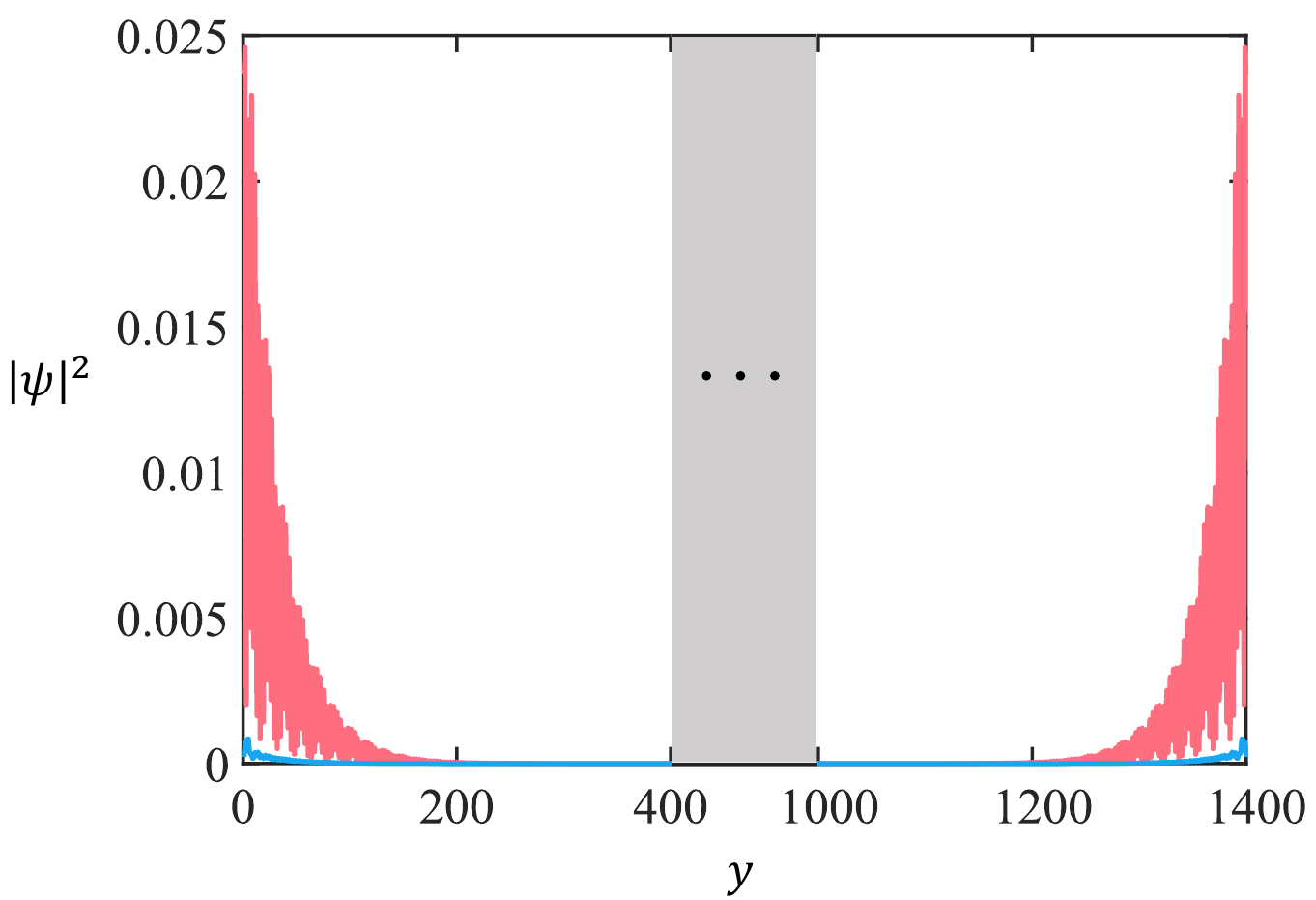}
    \caption{Real-space wave functions distribution of the zero-energy chiral edge modes corresponding to the energy crossing point at $k_x=0$. The red line represents the wave functions in the ferromagnetic metal layer, while the blue line corresponds to the wave functions in the superconductor layer. Parameters are taken the same as in Fig. 5(a).}
    \label{Fig:Psi}
\end{figure}

Since we are studying the two-layer system, we would like to examine the distribution of the wave functions of the edge states. In Fig. \ref{Fig:Psi}, we show the real-space distribution of the wave functions corresponding to the Chern number $C=-1$. It can be seen that the wave functions are mainly distributed at the edge of both the ferromagnetic metal layer and superconductor layer along the open boundary direction y. Moreover, the wave fuctions in this heterostructure are mainly concentrated in the ferromagnetic metal layer.

\section{conclusion}
In summary, we have studied a double-layer model consisting of spin-orbit coupling ferromagnetic metal and s-wave superconductor. The ferromagnetic metal layer has the intrinsic Dresselhaus spin-orbit coupling, while the two layers have the Rahsba spin-orbit interlayer hopping.
We calculated the Chern numbers of the system and demonstrated the phase diagrams. We found topological phases with five different Chern numbers. In particular, we found that the Chern number can be as large as four if the parameters of the systems are well controlled. We illustrated the Berry curvatures and showed that there are non-vanishing Berry curvature distributions in the momentum space even in the topologically trivial regime. We calculated the thermal Hall conductivity governed by the Berry curvatures. We revealed the chiral Majorana edge states protected by Chern numbers and by the winding numbers. We found that the wave functions of these edge states mostly distribute in the ferromagnetic metal layer.

\textit{Acknowledgments.---} We thank Zhongbo Yan, Jun He, and Dingyong Zhong for the valuable discussions. This project is supported by NKRDPC-2022YFA1402802, NKRDPC-2018YFA0306001, NSFC-92165204, NSFC-11974432, NSFC-12174453 and Shenzhen Institute for Quantum Science and Engineering.  

\bibliographystyle{apsrev4-2} 
\bibliography{Feynman}

\end{document}